\documentclass[11pt, a4paper,superscriptaddress]{article}
\usepackage{newtxtext,newtxmath}

\usepackage{graphicx}
\usepackage{parskip}
\usepackage[margin=1in]{geometry}
\usepackage[margin=1in]{geometry}
\usepackage{mathptmx} 
\usepackage{microtype}
\usepackage{setspace}
\usepackage{cite} 
\usepackage[colorlinks=true, linkcolor=blue, citecolor=blue, urlcolor=blue]{hyperref}

\title{\textbf{Reaction-induced Localized Annealing for On-chip Materials Discovery}}

\author{
  Nensi Toncich$^{1}$, Nerea Abando$^{1}$, Henning Galinski$^{1}$, Frans Spaepen$^{2}$ and Ralph Spolenak$^{1,*}$\\
  \small $^{1}$Laboratory for Nanometallurgy, Department of Materials, ETH Zurich, Zurich, 8093, Switzerland\\
  \small $^{2}$Harvard School of Engineering and Applied Sciences, Harvard University, Cambridge, MA 02138, USA\\
  \small $^{*}$Corresponding author email: \texttt{ralph.spolenak@mat.ethz.ch}
}
\date{}

\begin{document}
\newgeometry{top=0.8in, bottom=1in, left=1in, right=1in}
\maketitle
\begin{abstract}

On-chip thermal processing is a key technology for materials discovery across all disciplines of the natural sciences. Here, we present a microfabricated heat-delivery platform that enables reaction-induced localized on-chip annealing up to 1400~$^\circ$C. A buried reactive multilayer combined with a nanometric diffusion barrier provide ultra-fast heating and quenching rates on the order of $10^6~^\circ$C~s$^{-1}$ as illustrated by the generation of ultra-high aspect ratio microstructures during rapid-solidification and the stabilization of dopant-free tetragonal zirconia at room temperature. Rapid kinetics also enable the observation of transient states of capillary-driven hole collapse in melts. Our work expands localized high-temperature processing routes and offers a material-independent approach for rapid screening of multidimensional nanomaterials on and off-equilibrium. 
\end{abstract}

The net heat flux density refers to the rate of thermal energy that enters or leaves a material per unit area and sets the fundamental kinetic limit for atomic reorganization. Generating high net heat flux densities creates steep thermal gradients capable of trapping non-equilibrium states and amplifying phenomena, such as vitrification~\cite{Turnbull01091969} or the Soret effect~\cite{de2013non}.
\par
Traditionally, non-equilibrium states in materials, such as bainite~\cite{bhadeshia1980mechanism}, martensitic steel~\cite{gottstein2004physical}, amorphous silica~\cite{BRUCKNER1970123}, or metallic glasses~\cite{KLEMENT1960,johnson1999bulk,gallino2024physical}, have been realized using material specific processing routes that bypass the formation of equilibrium phases. Modern processing techniques, however, have drastically advanced the capabilities (Figure~\ref{fig:fig_1}A). Rapid thermal annealing (RTA) reduces dwell times in wafer-scale semiconductor manufacturing utilizing heating ramps on the order of 10-100~$^\circ$C$\cdot$s$^{-1}$, whereas flash lamp annealing (FLA)~\cite{Longfei2024} and flash Joule heating (FJH)~\cite{deng2025flash} compress thermal exposure to the millisecond-to-microsecond range making it ideal for thin-film recrystallization, dopant activation and defect annealing~\cite{rebohle2026crystallization}. Laser surface melting (LSM), electron-beam melting (EBM) and laser powder bed fusion (LPBF) combine high spatial control with high net heat flux densities reaching temporal thermal gradients in the range of $10^{4}$-$10^{6}~^\circ$C$\cdot$s$^{-1}$ and constitute key processes within additive manufacturing (AM)~\cite{wang2023understanding}.
\par
Still, beam-based heating processes rely on wave-matter interactions and require high-loss materials to localize energy effectively and enable steep thermal gradients. Methods based on Joule heating remain confined to materials whose resistance falls in a specific operational window, excluding most dielectrics and semiconductors. This high dependence on materials inherently limits their application as a universal heat delivery platform. 
\restoregeometry

\par
On-chip heat delivery offers a compelling alternative. Due to their characteristically low thermal mass, these integrated architectures enable steep heating and quenching rates. More importantly, a propagating thermal front converts heat delivery from a stationary exposure to a moving thermal pathway that can be engineered through the device architecture. As such, they represent a future gateway for efficient high-throughput synthesis~\cite{doi:10.1126/sciadv.aaz8867,doi:10.1126/sciadv.abg4930} and AI-driven materials-discovery~\cite{Mirkin2026}.  
\par 
Here, we present a reaction-induced localized on-chip annealing (RI-LOCA) platform that provides the foundation for such material-independent on-chip heat delivery.

\section*{On-chip heat platform design}

\begin{figure*} 
	\centering
	\includegraphics[width=0.99\textwidth]{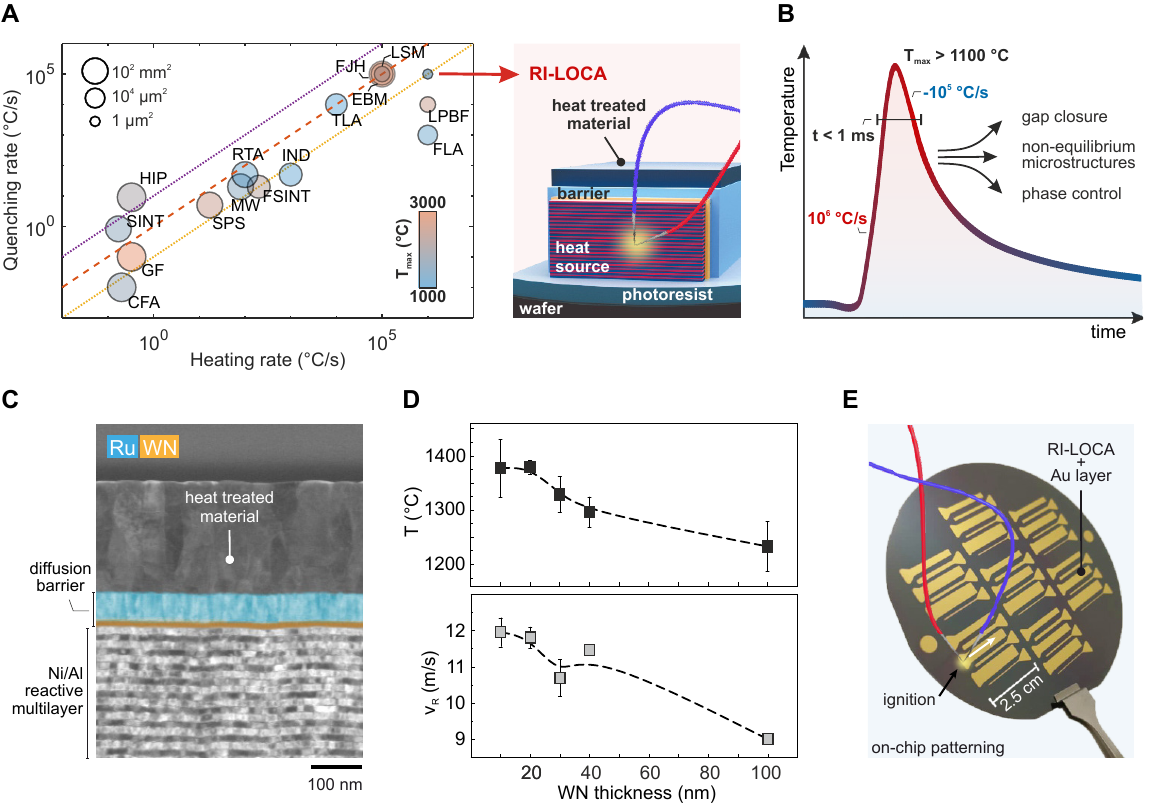}
\caption{\textbf{RI-LOCA enables localized, high-rate thermal processing on chip.}
(\textbf{A}) Thermal-processing landscape comparing established annealing routes by heating rate, quenching rate, processed area, and maximum temperature. Bubble size indicates characteristic processed area, and color indicates approximate maximum temperature. RI-LOCA occupies a high-rate, localized processing window; abbreviations are defined in the Supplementary Materials. The inset shows  a schematic cross-section of the RI-LOCA platform, with a reactive heat source separated from the heat-treated material by a diffusion barrier. 
(\textbf{B}) Schematic thermal profile of RI-LOCA, showing rapid heating above 1100$^\circ$C, a pulse duration below 1~ms, and fast cooling. This transient thermal pathway enables thin-film repair, nonequilibrium microstructure formation, and phase control. 
(\textbf{C}) Cross-sectional DF-STEM image of the RI-LOCA stack. The WN/Ru diffusion barrier separates the heat-treated material from the Ni/Al reactive multilayer; WN and Ru are highlighted in orange and light blue, respectively. 
(\textbf{D}) Peak temperature and reaction-front velocity, $v_R$, as a function of WN barrier thickness. Dashed lines are guides to the eye, and error bars indicate experimental uncertainty. 
(\textbf{E}) Optical image of patterned RI-LOCA devices on a Si chip with an Au overlayer. The heat-source geometry was defined using a steel shadow mask, confining the reaction and thermal pulse to selected on-chip regions. Electrical ignition of the RML heat source is initiated at the edge of the patterned structure using micromanipulator probes. The white arrow marks the resulting propagation direction of the heat front within the patterned RML region.}
	\label{fig:fig_1} 
\end{figure*}

The RI-LOCA platform uses a buried Ni/Al reactive multilayer (RML) as a self-propagating heat source, separated from the target material by a WN/Ru barrier (Figure~\ref{fig:fig_1}a,c). This architecture is designed to decouple the heat generation from the processed material. The exothermic solid-state reaction remains confined in the RML, while the heat is transferred through the barrier to the material above.
\par
Because the reactive zone reaches high temperatures and undergoes layer-specific melting during propagation, a perfect chemical isolation between the heat source and the to-be-heated material needs to be achieved. A nanometric WN/Ru bilayer provides the most effective barrier architecture identified in this work owing to the complementary properties of its two constituents (Figure~\ref{fig:fig_1}c). 
WN is a common low temperature diffusion-barrier material exhibiting high thermal stability and resistance to metal intermixing~\cite{kattelus1985sputtered,becker2003diffusion}. Ru provides a refractory, thermally conductive metallic layer that maintains efficient heat transfer and surface continuity during the short high-temperature pulse. Its structural stability at high temperature and large vacancy-formation and migration energies effectively suppress bulk lattice diffusion over the millisecond time scale of RI-LOCA processing~\cite{arblaster2013crystallographic,angsten2014elemental,daubriac2025overview}.
\par
With this WN/Ru barrier architecture, the chosen RI-LOCA configuration reaches a maximum surface temperature of 1377~$\pm$~54$^\circ$C and a maximum reaction-front velocity, $v_r$, of 12~$\pm$~0.4~m~s$^{-1}$. The corresponding heating and cooling rates are $3.3 \times 10^{6}$~$^\circ$C~s$^{-1}$ and $6.4 \times 10^{5}$~$^\circ$C~s$^{-1}$, respectively~\cite{toncich2025composition} (Figure~\ref{fig:fig_1}b). More generally, the heat release and propagation dynamics of Ni/Al reactive multilayers can also be tuned by changing the bilayer period, and composition~\cite{toncich2025composition,toncich2026ru}. Thus, the same device concept provides access to different localized heating and cooling profiles without changing the material placed on top.
\par
Patterning the reactive heat source confines the RI-LOCA thermal transient to predefined regions of the chip (Figure~\ref{fig:fig_1}e). Heat is generated exclusively where the patterned RML is ignited, enabling selective heating of thin-film libraries with limited heat transfer to surrounding areas. The lateral dimensions of the heat source can be controlled down to the sub-micrometer range, reaching feature sizes as small as 250~nm (see Supplementary Information).

\section*{Capillary-driven healing of a localized gap}

\begin{figure*} 
	\centering
	\includegraphics[width=0.99\linewidth]{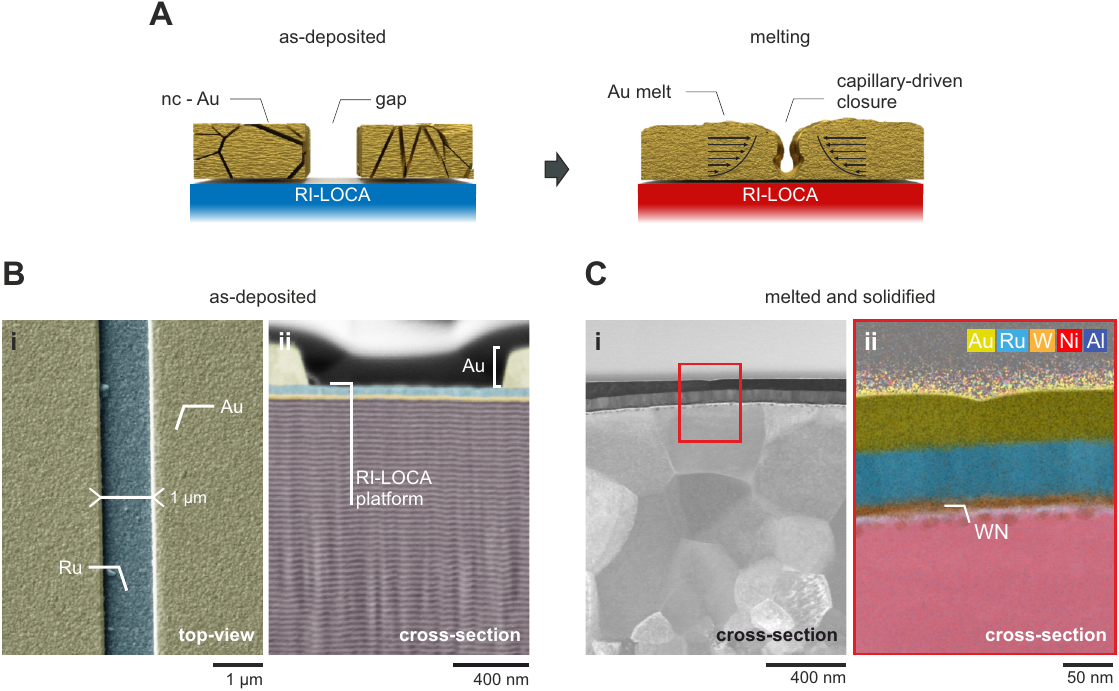} 

	\caption[RI-LOCA drives capillary-driven closure without intermixing.]{\textbf{RI-LOCA drives capillary-driven closure without intermixing.}
(\textbf{A}) Schematic of the gap-closure experiment. A nanocrystalline Au film containing a milled gap is deposited on the RI-LOCA platform. During ignition, the propagating reaction front locally melts the Au layer, and capillary forces drive molten Au from both sides toward the gap, promoting gap closure.
(\textbf{B}) As-deposited Au gap before RI-LOCA processing. (i) Color-coded top-view SEM image showing a 1~$\mu$m milled gap in the Au overlayer. (ii) Cross-sectional color-coded SEM image confirming that the gap is confined to the Au layer and that the underlying RI-LOCA platform remains intact.
(\textbf{C}) Cross-sectional analysis after RI-LOCA processing. (i) DF-STEM image showing local Au redistribution across the gap after annealing. The red box marks the region analyzed by elemental mapping. (ii) STEM-EDX map of the boxed region showing a continuous Au layer above the WN/Ru barrier and Ni/Al reactive multilayer. The preserved WN/Ru barrier suppresses detectable intermixing between the Au film and the reactive heat source.}
	\label{fig:fig_2} 
\end{figure*}

To test the performance of the heat delivery platform, we analyze the capillary-driven healing of a pre-fabricated gap in a molten gold thin film (Figure~\ref{fig:fig_2}a), a textbook problem of lubrication theory~\cite{Backholm2014}. A 200~nm thick Au layer was deposited by magnetron sputtering on top of the RI-LOCA stack, and artificial gaps up to 1 $\mu$m wide were milled only through the Au layer using focused ion beam etching~\cite{sm_example} (Figure~\ref{fig:fig_2}b).
\par
Upon heating above the melting temperature, the capillary pressure of the melt drives the leveling of the gap with a velocity that scales with the ratio of surface energy $\gamma(T)$ and viscosity $\mu(T)$ of the melt (see Supplementary Information).  Cross-sectional imaging after RI-LOCA processing confirms the melting of the gold film and the closure of the gap (Figure~\ref{fig:fig_2}c(i)). Importantly, the WN/Ru barrier remains preserved beneath the repaired region, including below the original FIB-milled gap. STEM-EDX mapping further shows no detectable intermixing of the Au layer and the elements of the heat delivery platform (Figure~\ref{fig:fig_2}c(ii)). 
\par
Using the height profile of the re-solidified film, we can access the thermal evolution in the vicinity of the gap and find that the healing of the gap is frozen in a transient state as the melt lifetime is only on the order of tens of microseconds (see Supplementary Information).

\section*{Rapid resolidification}

We leverage the steep thermal gradients of the RI-LOCA platform to probe the transient melting and re-solidification of gold thin films. Solidification provides an ideal testbed to study the emergence of non-equilibrium phenomena, once its velocity reaches a centimeter per second scale~\cite{https://doi.org/10.1002/1527-2648}. This transition can be studied by microscopy, as the final microstructure acts as a direct blueprint of the kinetics and thermodynamic forces at play.
\par
\begin{figure*}[h!] 
	\centering
	\includegraphics[width=1\textwidth]{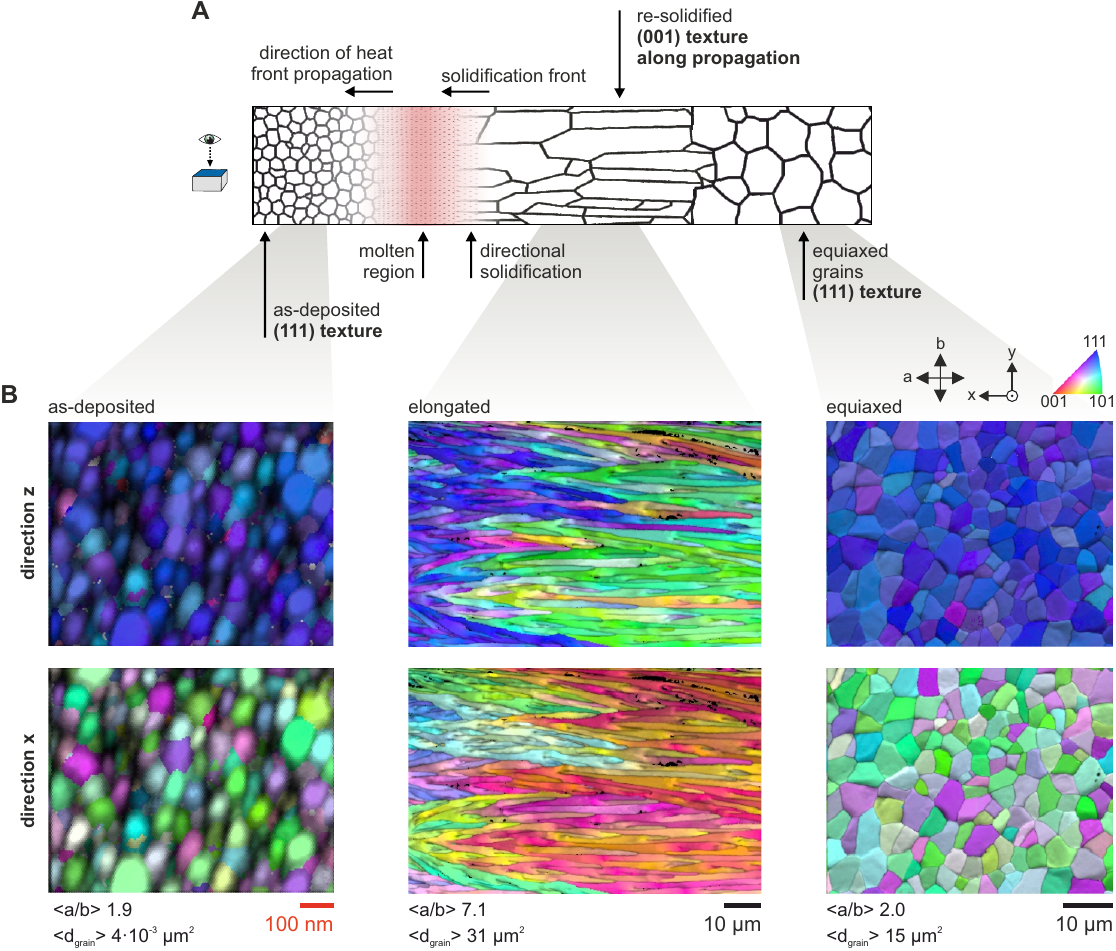}
	\caption{\textbf{RI-LOCA drives alternating directional and equiaxed re-solidification in Au thin films.}
(\textbf{A}) Schematic of the proposed re-solidification pathway during RI-LOCA processing. The propagating heat front locally melts the Au film and is followed by a solidification front. Behind this moving thermal field, the film re-solidifies into alternating regions of elongated grains, aligned with the propagation direction, and larger equiaxed grains.
(\textbf{B}) EBSD inverse pole figure maps of the Au film in the as-deposited state and after RI-LOCA processing, shown for the film-normal direction ($z$) and the in-plane propagation direction ($x$). Before processing, the Au film consists of fine, nearly equiaxed grains with a predominant out-of-plane (111) texture. After RI-LOCA, the microstructure alternates between elongated grains with near-(001) alignment along the propagation direction and larger equiaxed grains that preserve the as-deposited (111) out-of-plane texture. Mean aspect ratios and projected grain areas are indicated below each map.}
	\label{fig:fig_3} 
\end{figure*}

Before ignition, EBSD analysis of the Au film on the RI-LOCA platform shows a fine, nearly equiaxed as-deposited microstructure with a predominant out-of-plane $(111)$ texture (Figure~\ref{fig:fig_3}a,b (left)). Quantitative grain analysis gives a mean projected grain area of $4\times10^{-3}~\mu\mathrm{m}^2$ and a mean aspect ratio of 1.9. After propagation of the RI-LOCA front, the film develops an alternating microstructure of elongated and equiaxed grains with distinct textures (Figure~\ref{fig:fig_3}a,b (center,right)). The elongated grains align with the propagation direction, reach a mean projected area of $31~\mu\mathrm{m}^2$ and a mean aspect ratio of 7.1, and show a near-$(001)$ alignment along the direction of elongation. By contrast, the equiaxed regions preserve the as-deposited $(111)$ out-of-plane texture and consist of larger, nearly equiaxed grains with a mean aspect ratio of 2.0 and a mean projected grain area of $15~\mu\mathrm{m}^2$.
\par
The formation of this stop-and-go microstructure indicates that the local solidification regime changes during front propagation. A possible origin is unsteady propagation of the reactive multilayer front, since RMLs can exhibit pulsating reaction fronts that generate spatially modulated temperature fields under specific material, interfacial, and thermal-loss conditions~\cite{adams2015reactive,abere2023effects,kim2025numerical}. Such local variations could shift the Au film between complete melting followed by directional re-solidification and partial melting followed by seeded equiaxed regrowth.\par
Because RI-LOCA heats the Au film to its melting threshold, slightly lower local thermal exposure may leave residual crystalline cores. In nanocrystalline Au, melting is expected to initiate preferentially at grain boundaries and propagate into grain interiors~\cite{liu2018melting,assefa2020ultrafast}. These residual cores could seed rapid regrowth during cooling, producing large equiaxed grains while preserving part of the as-deposited texture.\par
In fully molten regions, by contrast, orientation memory is erased and grains re-solidify directionally behind the moving thermal field. For FCC metals, rapid solidification commonly favors growth along $\langle 100\rangle$-type directions. Grains with a $[001]$ direction aligned with the local temperature gradient therefore grow preferentially and overgrow less favorably oriented neighbors~\cite{debroy2018additive,kulovits2009electron}. This directional selection explains the elongated grains and their near-$[001]$ alignment along the propagation direction.

\section*{Kinetic phase stabilization}

\begin{figure*}[h!] 
	\centering
	\includegraphics[width=1\textwidth]{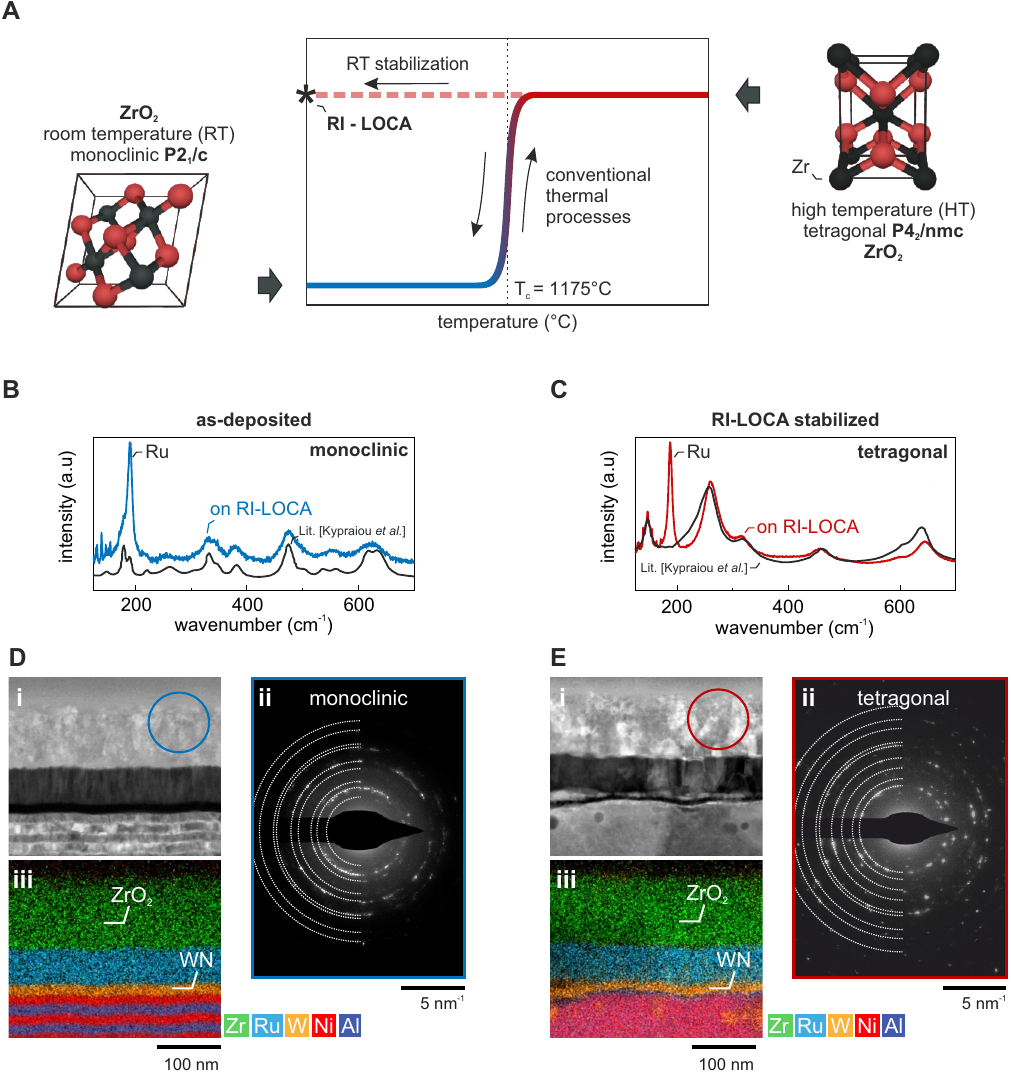} 

	\caption{\textbf{RI-LOCA stabilizes dopant-free tetragonal ZrO$_2$.}
(\textbf{A}) Schematic of the ZrO$_2$ phase-control concept. Undoped ZrO$_2$ is monoclinic at room temperature, whereas the tetragonal phase is stable at high temperature. Conventional thermal processing allows back-transformation during cooling, while RI-LOCA provides a short high-temperature excursion followed by rapid quenching, enabling room-temperature retention of the tetragonal phase.
(\textbf{B}) Raman spectrum of the as-deposited ZrO$_2$ layer on the RI-LOCA platform. The measured spectrum is consistent with monoclinic ZrO$_2$ reference features~\cite{kypraiou2012identification}; the Ru peak originates from the underlying Ru layer.
(\textbf{C}) Raman spectrum after RI-LOCA processing. The phase-sensitive ZrO$_2$ bands shift to a tetragonal signature~\cite{kypraiou2012identification}, indicating stabilization of tetragonal ZrO$_2$ after the localized thermal pulse.
(\textbf{D}) Structural and chemical analysis of the as-deposited state. (\textbf{i}) Cross-sectional DF-STEM image of the ZrO$_2$/RI-LOCA stack. (\textbf{ii}) Selected-area electron diffraction pattern aquired from the topmost ZrO$_2$ layer and indexed to monoclinic ZrO$_2$. (\textbf{iii}) STEM-EDX elemental map showing the ZrO$_2$, WN/Ru barrier, and Ni/Al reactive multilayer sequence.
(\textbf{E}) Structural and chemical analysis after RI-LOCA processing. (\textbf{i}) Cross-sectional DF-STEM image of the processed stack. (\textbf{ii}) Selected-area electron diffraction pattern aquired from the topmost ZrO$_2$ layer and indexed to tetragonal ZrO$_2$. (\textbf{iii}) STEM-EDX elemental map showing preservation of the layered architecture after ignition.}
	\label{fig:fig_4} 
\end{figure*}

To showcase the versatility of the heat delivery platform, we demonstrate suppression of the tetragonal-to-monoclinic inversion in $100$~nm thick zirconia films. ZrO$_2$ is monoclinic at room temperature, while the tetragonal phase exists only at temperatures between $1175~^\circ$C and $2370~^\circ$C~\cite{kingery1976introduction} (Figure~\ref{fig:fig_4}a). In bulk materials, stabilization of the tetragonal phase is achieved exclusively by altering the free energy and oxygen-vacancy population via doping~\cite{shukla2005mechanisms}. On the nanoscale, the stabilization of pure ZrO$_2$ has been demonstrated only in nanoparticle systems with a critical particle size below 30~nm~\cite{garvie1978stabilization,shukla2005mechanisms}. So far, no consensus on the underlying mechanism that drives this transformation has been achieved~\cite{shukla2005mechanisms}. 
\par
Monoclinic ZrO$_2$ films were deposited by reactive-sputtering onto the WN/Ru-protected Ni/Al RI-LOCA platform. Structural and chemical analysis, including Raman spectroscopy, cross-sectional transmission electron microscopy (TEM) and selected-area electron diffraction (SAED) indicate that these films are nanocrystalline with an average grain size of 3-7~nm (Figure~\ref{fig:fig_4}b,d). The SAED pattern and Raman spectrum were successfully indexed to the monoclinic crystal structure (Figure~\ref{fig:fig_4}b,d(i,ii)), while STEM-EDX elemental mapping and Rutherford backscattering (see Supplementary Information) spectrometry were employed to guarantee high purity and exact stoichiometry of the film.
\par
Following a rapid heating and quenching cycle on the RI-LOCA platform, the crystallographic signature of the film shifts significantly. The resulting Raman spectrum and SAED pattern match the characteristic response of tetragonal ZrO$_2$, confirming the quenching of the high temperature phase (Figure~\ref{fig:fig_4}c,e(i,ii)). STEM images show that the ZrO$_2$ film remains supported on an intact WN/Ru barrier and an underlying reacted Ni/Al RML after RI-LOCA processing (Figure~\ref{fig:fig_4}e(iii)).
\par
The stabilization of the tetragonal phase cannot be understood in the context of a classical displacive martensitic transformation which propagates with the speed of sound, vastly exceeding the quenching rate imposed by RI-LOCA~\cite{kingery1976introduction,chiao1990martensitic}. Retention of tetragonal ZrO$_2$ after processing therefore suggests that, in this supported thin-film geometry, the reverse transformation is not governed solely by the intrinsic shear step, but also by the conditions required for nucleation, accommodation, and completion of the transformation. The present ex-situ measurements do not directly resolve the atomic pathway during the RI-LOCA transient, and therefore cannot determine whether the transformation remains purely diffusionless under these non-equilibrium thin-film conditions or whether additional defect-, interface-, or relaxation-mediated processes contribute. Establishing that distinction would require time-resolved structural probes during the thermal transient, or atomically resolved measurements immediately after processing.
  
\section*{Concluding remarks}

RI-LOCA establishes a local, high-rate route to thin-film thermal processing on chip. A patterned Ni/Al reactive multilayer generates a propagating thermal front in predefined regions, while a WN/Ru barrier separates the reaction from the material being processed. This architecture enables tunable high-temperature pulses without global annealing and can be applied to different surface-supported materials.
\par
Using Au and ZrO$_2$ as model systems, we show that a single confined thermal event can drive capillary-mediated closure of Au gaps, produce alternating high-aspect-ratio and equiaxed Au grain regions with distinct crystallographic textures, and retain dopant-free tetragonal ZrO$_2$ at room temperature.
\par
By localizing heat generation down to sub-micrometer scales, RI-LOCA establishes a powerful platform for spatial thermal processing of thin-film material libraries in or out of equilibrium. When coupled with high-throughput material synthesis and AI-driven materials discovery, this spatial control can open unprecedented opportunities to engineer and map metastable phases, texture states, and broken-symmetry phenomena across complex oxides, two-dimensional systems, and chiral architectures.



%


\section*{Acknowledgments}
We thank Zhichao Chen, Jacopo Crippa, Jemma Gillon, Patricia Moll, and Kristian Skailand for their contributions to the systematic screening and optimization of diffusion barriers for the RI-LOCA platform. We thank Marcello Pozzi for the RBS measurement and Judith Gómez Cuyàs for the Raman measurements. We gratefully acknowledge access to the electron microscopy instrumentation at ScopeM, the microscopy platform of ETH Zürich, and to the sputter-deposition facilities at FIRST, the cleanroom facility of ETH Zürich.\par
\scriptsize
\textbf{Funding:} No specific funding was received for this work.\par
\textbf{Author contributions:} Conceptualization: N.T., H.G., R.S.; Methodology: N.T., H.G., R.S.; Investigation: N.T., N.A.; Formal analysis: N.T., N.A., H.G.; Data interpretation and discussion: N.T., N.A., H.G., F.S., R.S.; Writing - original draft: N.T., H.G; Writing - review \& editing: N.T., N.A., H.G., F.S., R.S.; Visualization: N.T., N.A., H.G.; Supervision: R.S.; Project administration: R.S.\par
\textbf{Competing interests:} The authors declare no competing interests.\par
\textbf{Data, code, and materials availability:} All data needed to evaluate the conclusions of this study are present in the paper and the Supplementary Materials.

\begin{thebibliography}{10}
\providecommand{\url}[1]{\texttt{#1}}
\expandafter\ifx\csname urlstyle\endcsname\relax
  \providecommand{\doi}[1]{doi:\discretionary{}{}{}#1}\else
  \providecommand{\doi}{doi:\discretionary{}{}{}\begingroup \urlstyle{rm}\Url}\fi

\bibitem{Turnbull01091969}
D.~Turnbull, Under what conditions can a glass be formed? \emph{Contemporary Physics} \textbf{10}~(5), 473--488 (1969).

\bibitem{de2013non}
S.~R. De~Groot, P.~Mazur, \emph{Non-equilibrium thermodynamics} (Courier Corporation) (2013).

\bibitem{bhadeshia1980mechanism}
H.~Bhadeshia, D.~Edmonds, The mechanism of bainite formation in steels. \emph{Acta Metallurgica} \textbf{28}~(9), 1265--1273 (1980).

\bibitem{gottstein2004physical}
G.~Gottstein, \emph{Physical foundations of materials science}, vol.~3 (Springer) (2004).

\bibitem{BRUCKNER1970123}
Properties and structure of vitreous silica. I. \emph{Journal of Non-Crystalline Solids} \textbf{5}~(2), 123--175 (1970).

\bibitem{KLEMENT1960}
W.~Klement, R.~H. Willens, P.~O.~L. Duwez, Non-crystalline Structure in Solidified Gold-Silicon Alloys. \emph{Nature} \textbf{187}~(4740), 869--870 (1960).

\bibitem{johnson1999bulk}
W.~L. Johnson, Bulk glass-forming metallic alloys: Science and technology. \emph{MRS bulletin} \textbf{24}~(10), 42--56 (1999).

\bibitem{gallino2024physical}
I.~Gallino, R.~Busch, \emph{Physical metallurgy of bulk metallic glass-forming liquids} (Springer) (2024).

\bibitem{Longfei2024}
L.~Song, \emph{et~al.}, Crystallization of piezoceramic films on glass via flash lamp annealing. \emph{Nature Communications} \textbf{15}~(1), 1890-- (2024).

\bibitem{deng2025flash}
B.~Deng, L.~Eddy, K.~M. Wyss, C.~S. Tiwary, J.~M. Tour, Flash Joule heating for synthesis, upcycling and remediation. \emph{Nature Reviews Clean Technology} \textbf{1}~(1), 32--54 (2025).

\bibitem{rebohle2026crystallization}
L.~Rebohle, \emph{et~al.}, Crystallization of Semiconductor Thin Films by Flash Lamp Annealing. \emph{physica status solidi (a)} \textbf{223}~(7), e202600006 (2026).

\bibitem{wang2023understanding}
J.~Wang, R.~Zhu, Y.~Liu, L.~Zhang, Understanding melt pool characteristics in laser powder bed fusion: An overview of single-and multi-track melt pools for process optimization. \emph{Advanced Powder Materials} \textbf{2}~(4), 100137 (2023).

\bibitem{doi:10.1126/sciadv.aaz8867}
B.~P. MacLeod, \emph{et~al.}, Self-driving laboratory for accelerated discovery of thin-film materials. \emph{Science Advances} \textbf{6}~(20), eaaz8867 (2020).

\bibitem{doi:10.1126/sciadv.abg4930}
S.~Ament, \emph{et~al.}, Autonomous materials synthesis via hierarchical active learning of nonequilibrium phase diagrams. \emph{Science Advances} \textbf{7}~(51), eabg4930 (2021).

\bibitem{Mirkin2026}
J.~Li, \emph{et~al.}, High-entropy 1D halide perovskite piezoelectrics found by megalibrary synthesis and rapid nonlinear optical screening. \emph{Science Advances} \textbf{12}~(21), eaee1359 (2026).

\bibitem{kattelus1985sputtered}
H.~Kattelus, E.~Kolawa, K.~Affolter, M.-A. Nicolet, Sputtered W--N diffusion barriers. \emph{Journal of Vacuum Science \& Technology A: Vacuum, Surfaces, and Films} \textbf{3}~(6), 2246--2254 (1985).

\bibitem{becker2003diffusion}
J.~S. Becker, R.~G. Gordon, Diffusion barrier properties of tungsten nitride films grown by atomic layer deposition from bis (tert-butylimido) bis (dimethylamido) tungsten and ammonia. \emph{Applied Physics Letters} \textbf{82}~(14), 2239--2241 (2003).

\bibitem{arblaster2013crystallographic}
J.~W. Arblaster, Crystallographic properties of ruthenium. \emph{Platinum Metals Review} \textbf{57}~(2), 127--136 (2013).

\bibitem{angsten2014elemental}
T.~Angsten, T.~Mayeshiba, H.~Wu, D.~Morgan, Elemental vacancy diffusion database from high-throughput first-principles calculations for fcc and hcp structures. \emph{New Journal of Physics} \textbf{16}~(1), 015018 (2014).

\bibitem{daubriac2025overview}
R.~Daubriac, \emph{et~al.}, Overview of Ruthenium Thin Films Annealed by Microsecond Scanning UV Pulsed Laser: Structural, Electrical, and Failure Modes Analysis. \emph{Advanced Engineering Materials} \textbf{27}~(4), 2402656 (2025).

\bibitem{toncich2025composition}
N.~Toncich, F.~Schwarz, R.~A. Gallivan, J.~Gillon, R.~Spolenak, Composition effects on Ni/Al reactive multilayers: A comprehensive study of mechanical properties, reaction dynamics, and phase evolution. \emph{Journal of Applied Physics} \textbf{137}~(17) (2025).

\bibitem{toncich2026ru}
N.~Toncich, A.~Yadav, J.~Fikar, R.~Spolenak, Ru alloying in Ni/Al reactive multilayers: Experimental observations and molecular dynamics simulations. \emph{Journal of Applied Physics} \textbf{140}~(2), 025102 (2026).

\bibitem{Backholm2014}
M.~Backholm, M.~Benzaquen, T.~Salez, E.~Raphaël, K.~Dalnoki-Veress, Capillary levelling of a cylindrical hole in a viscous film. \emph{Soft Matter} \textbf{10}~(15), 2550--2558 (2014).

\bibitem{sm_example}
\emph{See the supplementay materials}.

\bibitem{https://doi.org/10.1002/1527-2648}
W.~Kurz, Solidification Microstructure-Processing Maps: Theory and Application. \emph{Advanced Engineering Materials} \textbf{3}~(7), 443--452.

\bibitem{adams2015reactive}
D.~Adams, Reactive multilayers fabricated by vapor deposition: A critical review. \emph{Thin Solid Films} \textbf{576}, 98--128 (2015).

\bibitem{abere2023effects}
M.~J. Abere, \emph{et~al.}, Effects of diffusion barriers on reaction wave stability in Co/Al reactive multilayers. \emph{Journal of Applied Physics} \textbf{134}~(19), 195302 (2023).

\bibitem{kim2025numerical}
K.~Kim, Numerical simulation on periodic advancing of reaction front in reactive multilayer nanolaminates. \emph{Journal of Mechanical Science and Technology} pp. 1--9 (2025).

\bibitem{liu2018melting}
J.~Liu, X.~Fan, Y.~Shi, D.~J. Singh, W.~Zheng, Melting of nanocrystalline gold. \emph{The Journal of Physical Chemistry C} \textbf{123}~(1), 907--914 (2018).

\bibitem{assefa2020ultrafast}
T.~A. Assefa, \emph{et~al.}, Ultrafast x-ray diffraction study of melt-front dynamics in polycrystalline thin films. \emph{Science advances} \textbf{6}~(3), eaax2445 (2020).

\bibitem{debroy2018additive}
T.~DebRoy, \emph{et~al.}, Additive manufacturing of metallic components--process, structure and properties. \emph{Progress in materials science} \textbf{92}, 112--224 (2018).

\bibitem{kulovits2009electron}
A.~Kulovits, R.~Zhong, J.~Wiezorek, J.~Leonard, Electron microscopy of geometrically confined copper thin films after rapid lateral solidification. \emph{Thin Solid Films} \textbf{517}~(13), 3629--3634 (2009).

\bibitem{kingery1976introduction}
W.~D. Kingery, H.~K. Bowen, D.~R. Uhlmann, \emph{Introduction to ceramics} (John wiley \& sons) (1976).

\bibitem{shukla2005mechanisms}
S.~Shukla, S.~Seal, Mechanisms of room temperature metastable tetragonal phase stabilisation in zirconia. \emph{International materials reviews} \textbf{50}~(1), 45--64 (2005).

\bibitem{garvie1978stabilization}
R.~Garvie, Stabilization of the tetragonal structure in zirconia microcrystals. \emph{The Journal of Physical Chemistry} \textbf{82}~(2), 218--224 (1978).

\bibitem{chiao1990martensitic}
Y.~Chiao, I.-W. Chen, Martensitic growth in ZrO2—An in situ, small particle, TEM study of a single-interface transformation. \emph{Acta Metallurgica et Materialia} \textbf{38}~(6), 1163--1174 (1990).

\bibitem{kypraiou2012identification}
V.~Kypraiou, S.~Pelekanos, G.~Eliades, Identification of monoclinic phase in CAD/CAM zirconia FPD frameworks. \emph{Eur J Esthet Dent} \textbf{7}~(4), 418--429 (2012).

\end{thebibliography}
\end{document}